\documentclass[english,aip,jcp,english,dvipsnames]{revtex4-1}
\usepackage[T1]{fontenc}
\usepackage[latin9]{inputenc}
\usepackage{float}
\usepackage{textcomp}
\usepackage{amsmath}
\usepackage{amssymb}
\usepackage{graphicx}
\usepackage{esint}
\usepackage{subscript}

\makeatletter

\providecommand{\tabularnewline}{\\}

 \@ifundefined{textcolor}{}
 {%
   \definecolor{BLACK}{gray}{0}
   \definecolor{WHITE}{gray}{1}
   \definecolor{RED}{rgb}{1,0,0}
   \definecolor{GREEN}{rgb}{0,1,0}
   \definecolor{BLUE}{rgb}{0,0,1}
   \definecolor{CYAN}{cmyk}{1,0,0,0}
   \definecolor{MAGENTA}{cmyk}{0,1,0,0}
   \definecolor{YELLOW}{cmyk}{0,0,1,0}
 }

\usepackage{setspace}
\renewcommand{\textendash}{--}
\renewcommand{\textemdash}{---}

\makeatother

\usepackage{babel}
\begin{document}

\title{Semiclassical Vibrational Spectroscopy with Hessian Databases}

\author{Riccardo Conte}
\email{riccardo.conte1@unimi.it}

\affiliation{Dipartimento di Chimica, Università degli Studi di Milano, via Golgi
19, 20133 Milano, Italy.}

\author{Fabio Gabas}

\affiliation{Dipartimento di Chimica, Università degli Studi di Milano, via Golgi
19, 20133 Milano, Italy.}

\author{Giacomo Botti}

\affiliation{Dipartimento di Chimica, Università degli Studi di Milano, via Golgi
19, 20133 Milano, Italy.}

\author{Yu Zhuang}
\email{yu.zhuang@ttu.edu}

\affiliation{Department of Computer Science, Texas Tech University, Lubbock, Texas
79409-3104, USA.}

\author{Michele Ceotto}
\email{michele.ceotto@unimi.it}

\affiliation{Dipartimento di Chimica, Università degli Studi di Milano, via Golgi
19, 20133 Milano, Italy.}
\begin{abstract}
We report on a new approach to ease the computational overhead of
ab initio ``on-the-fly'' semiclassical dynamics simulations for
vibrational spectroscopy. The well known bottleneck of such computations
lies in the necessity to estimate the Hessian matrix for propagating
the semiclassical pre-exponential factor at each step along the dynamics.
The procedure proposed here is based on the creation of a dynamical
database of Hessians and associated molecular geometries able to speed
up calculations while preserving the accuracy of results at a satisfactory
level. This new approach can be interfaced to both analytical potential
energy surfaces and on-the-fly dynamics, allowing one to study even
large systems previously not achievable. We present results obtained
for semiclassical vibrational power spectra of methane, glycine, and
N-acetyl-L-phenylalaninyl-L-methionine-amide, a molecule of biological
interest made of 46 atoms.
\end{abstract}
\maketitle

\section{Introduction}

Several computational methods for calculating quantum molecular vibrational
spectra have been established. To name some popular ones, we recall
the vibrational self-consistent field and virtual space configuration
interaction (VSCF/VCI) method,\citep{Bowman_Huang_MULTIMODE_2003,Bowman_Meyer_Polyatomic_2008,huang_bowman_RPMDzundel_2008,mancini_Bowman_mizedwaterHClclusters_2014}
the multi-configuration time dependent Hartree (MCTDH) approach,\citep{Meyer_Cederbaum_MCTDH_1990,Meyer_Worth_HighDimMCTDH_2003,vendrell_Meyer_zundelspectra_2007,Picconi_Burghardt_I2Kr18_2019,Picconi_Burghardt_I2Kr18_II_2019}
the use of contracted basis functions with an eigensolver,\citep{Cullum_Willoughby_LanczosDiag_1985,davidson_Eigensolver_1975}
collocation methods,\citep{Manzhos_Carrington_RectangularCollocation_2016,Avila_Carrington_SmolyakCollocation_2017}
and second-order vibrational perturbation theory (VPT2).\citep{Barone_AutomatedVPT2_2005}
Despite the advance these methods provide in dealing with the exponential
scaling of quantum approaches with the number of vibrational degrees
of freedom, the investigation of large molecules or complex supra-molecular
systems remains quite challenging. In this case classical simulations
or ad-hoc scaling of harmonic estimates are the approaches commonly
employed in spite of the drawback of totally neglecting or not properly
describing quantum effects (such as zero-point energies, quantum couplings,
and overtones). This aspect has been recently confirmed by a semiclassical
(SC) investigation,\citep{Gabas_Ceotto_SupramolecularGlycines_2018}
which has explained experimental findings due to quantum effects,
in contrast to previous non conclusive theoretical attempts.\citep{masson_Rizzo_GlycineH2_2015,oh_mclafferthy_electrospray_2005}
Furthermore, recent SC techniques which are capable of accurately
calculating the intensities of vibrational spectral transitions have
demonstrated that the widely employed double harmonic approximation
often provides only a rough estimate, which has to be corrected by
incorporating quantum couplings and anharmonicities.\citep{Micciarelli_Ceotto_SCwavefunctions_2018,Micciarelli_Ceotto_SCwavefunctions2_2019} 

Semiclassical dynamics is indeed gaining more and more attention due
to some of its peculiar features, which include the possibility of
accounting for quantum effects in an accurate way starting from short-time
classical trajectories,\citep{Miller_QTST_1974,Heller_FrozenGaussian_1981,Herman_Kluk_SCnonspreading_1984,Miller_SCnorootsearch_1991,Kay_Multidim_1994,Walton_Manolopoulos_FrozenGaussianCO2_1995,Elran_Kay_ImprovingHK_1999,Grossmann_HierarchySC_1999,Shalashilin_Child_Coherentstates_2001,Miller_PNAScomplexsystems_2005,Kay_Atomsandmolecules_2005,Liu_Miller_linearizedSCIVR_2007,Ceotto_AspuruGuzik_Multiplecoherent_2009,Ceotto_Tantardini_Copper100_2010,Ceotto_AspuruGuzik_Curseofdimensionality_2011,Tamascelli_Ceotto_GPU_2014,Wehrle_Vanicek_NH3_2015,Ceotto_Buchholz_SAM_2018,Ma_Ceotto_SN2reactions_2018,Conte_Ceotto_book_chapter_2019,Church_Ananth_Filinov_2017,Bonella_Coker_Linearizedpathintegral_2005,Tao_Miller_Tdepsampling_2011,Tao_Miller_Tdepsampling_2012,Tao_ImportanceSampling_2014,Tao_Nonadiabatic_2013}
the advantage of not being limited to a single-well picture,\citep{Conte_Ceotto_NH3_2013,Gabas_Ceotto_Glycine_2017,Ceotto_watercluster_18}
and the possibility of adopting an ab initio ``on-the-fly'' approach.\citep{Ceotto_AspuruGuzik_PCCPFirstprinciples_2009,Tatchen_Pollak_Onthefly_2009,Ceotto_AspuruGuzik_Firstprinciples_2011,Wehrle_Vanicek_Oligothiophenes_2014,Patoz_Vanicek_Photoabs_Benzene_2018,Gabas_Ceotto_Glycine_2017}
These characteristics make the range of applicability of semiclassical
spectroscopy enormously vast, ranging from small molecules, for which
very precise analytical potential energy surfaces (PES) can be constructed,
to larger molecules or even supra-molecular systems whose dynamics
can be integrated ``on-the-fly'' from ab initio calculations. Clearly,
a very different computational effort is required depending on whether
calls to the potential are analytical or ask for ab initio electronic
structure calculations as in the case of ``on-the-fly'' simulations.\citep{hase_pratihar_directdynamics_2017,Marx_Hutter_2009book,marx1996ab,Gaigeot_Castiglioni_IRclusters,gaigeot_gaigeot_floppypeptides_2010}
In SC dynamics this difference translates mainly into the very different
number of trajectories on which SC calculations are based. In fact,
when working with an analytical PES and exploiting time-average (TA)
filtering techniques, a proper Monte Carlo convergence in the evaluation
of the semiclassical phase-space integration can be reached.\citep{Kaledin_Miller_Timeaveraging_2003,Kaledin_Miller_TAmolecules_2003}
A tailored approach, known as Multiple Coherent Semiclassical Initial
Value Representation (MC SCIVR), must be employed instead when adopting
``on-the-fly'' dynamics, so that even a single tailored trajectory
allows one to regain the main spectral features.\citep{Ceotto_AspuruGuzik_Multiplecoherent_2009,Gabas_Ceotto_Glycine_2017} 

Application of standard SC techniques is hampered by a few well-known
drawbacks. One is the so-called ``curse of dimensionality'' which
concerns the impossibility of obtaining a resolved spectroscopic signal
when the dimensionality of the system becomes too large. Another one
is related to the chaotic behavior of classical dynamics. Chaotic
trajectories make numerical integration of the monodromy matrix elements
(needed to evaluate the SC frozen Gaussian propagator) inaccurate
because of finite machine precision. The consequence is that, if the
contribution of these trajectories is neither discarded nor reshaped,
the entire calculation may be spoiled. However, even in the case of
non-chaotic trajectories, propagation of the monodromy matrix elements
represents a computational bottleneck due to the necessity to evaluate
the Hessian matrix (matrix of second derivatives of the potential
with respect to the nuclear coordinates) at each step along the dynamics. 

Recent advances in the semiclassical field have permitted researchers
to make progress on these issues and further research is still being
carried out. For instance, a ``divide-and-conquer'' strategy (DC
SCIVR) has been employed to extend semiclassical spectroscopy to systems
as large as fullerene;\citep{ceotto_conte_DCSCIVR_2017,DiLiberto_Ceotto_Jacobiano_2018,Ceotto_watercluster_18}
appropriate approximations have been introduced to deal with chaotic
trajectories;\citep{Kay_Numerical_1994,Gelabert_Miller_logderivative_2000,DiLiberto_Ceotto_Prefactors_2016,Gabas_Ceotto_Glycine_2017}
and a compact finite-difference (CFD) approximation to the Hessian
has been able to reduce the number of Hessian calculations required.\citep{Wu_Zhuang_HessApprox_2010,Zhuang_Ceotto_Hessianapprox_2012,Ahmadian_Chen_BigDataZhuang_2017}
In particular, CFD-Bofill schemes permit one to preserve the accuracy
in the propagation of monodromy matrix elements quite well and, using
them, accurate semiclassical spectra have been obtained for several
small molecules including an ``on-the-fly'' application to CO\textsubscript{2}.\citep{Ceotto_Hase_AcceleratedSC_2013} 

So far this introduction has been focused on SC frozen Gaussian propagators,
which were employed in the numerical applications presented in this
paper. However, Hessian calculations are ubiquitous in semiclassical
dynamics and, in fact, they are required also by another variety of
SC propagators, known as thawed Gaussian propagators.\citep{Heller_TdependentSC_1975,Baranger_Schellhaass_2001,Pollak_Artes_ThawedGaussian_2004,Grossmann_SChybrid_2006,Conte_Pollak_ThawedGaussian_2010}
The main difference between the two families is that in a frozen Gaussian
the width of the Gaussians is a constant in time, while in a thawed
Gaussian the width evolves in time. Furthermore, SC thawed Gaussian
calculations are based on single-trajectory dynamics and are known
to be less accurate than their frozen Gaussian counterparts even in
the case of MC SCIVR single trajectory simulations. Interpolation
schemes have been developed for thawed Gaussian propagators,\citep{Wehrle_Vanicek_Oligothiophenes_2014,Buchholz_Ceotto_MixedSC_2016,Ceotto_Buchholz_MixedSC_2017,Buchholz_Jungwirth_CondensedPhase_2012,Begusic_Vanicek_Electronicspectra_2018,Patoz_Vanicek_Photoabs_Benzene_2018}
and, at the time of writing, an approach based on a single, initial
Hessian evaluation has just been proposed and applied to SC thawed
Gaussian vibro-electronic (vibronic) calculations.\citep{Begusic_Vanicek_SingleHessian_2019} 

The main goals of this paper are the introduction of a new strategy
to slash the computational overhead associated with Hessian matrix
calculations and its application to precise semiclassical frozen Gaussian
vibrational spectroscopy of polyatomic molecular systems. The approach
is based on the dynamical construction of a database of Hessian matrices
dependent on the molecular geometry visited along the dynamics. It
is not as computationally inexpensive as the thawed Gaussian single
Hessian scheme, but more accurate. We show that this Hessian approximation
is accurate, and able to extend the range of applicability of ``on-the-fly''
SC vibrational spectroscopy to larger chemical species. The paper
introduces the theoretical details of the new approach in Section
\ref{sec:Theory}. Section \ref{sec:Results} is devoted to applications
to methane, glycine and N-acetyl-L-phenylalaninyl-L-methionine-amide.
Finally, in Section \ref{sec:Summary}, a summary and some conclusions
are reported.

\section{Theoretical and Computational Details\label{sec:Theory}}

The semiclassical calculation of vibrational power spectra of molecular
species is a long established technique, which does not suffer from
zero point energy (ZPE) leakage\citep{Buchholz_Ceotto_leakage_2018}
and has recently evolved into a very powerful computational tool able
to deal with systems involving dozens of degrees of freedom. In this
Section we introduce the SC theoretical methodology on which our results
are based, and then focus on the description of the new database strategy.

In a pivotal paper\citep{Kaledin_Miller_Timeaveraging_2003} Kaledin
and Miller demonstrated that by time averaging the Herman-Kluk semiclassical
propagator, it is possible to work out a computationally advantageous
expression for the power spectrum of a survival amplitude evolved
according to a vibrational Hamiltonian. Their starting point, the
Herman-Kluk (HK) propagator, provides a semiclassical approximation
to the exact quantum propagator. In particular, for an $N_{vib}$-dimensional
system, the survival amplitude of a generic reference state $|\chi\rangle$
can be written as

\begin{equation}
\left\langle \chi\left|e^{-i\hat{H}t/\hbar}\right|\chi\right\rangle _{HK}=\left(\frac{1}{2\pi\hbar}\right)^{N_{vib}}\iintop d\mathbf{p}_{0}d\mathbf{q}_{0}C_{t}\left(\mathbf{p}_{0},\mathbf{q}_{0}\right)e^{\frac{i}{\hbar}S_{t}\left(\mathbf{p}_{0},\mathbf{q}_{0}\right)}\left\langle \chi\right.\left|\mathbf{p}_{t},\mathbf{q}_{t}\left\rangle \right\langle \mathbf{p}_{0},\mathbf{q}_{0}\right|\left.\chi\right\rangle .\label{eq:HHKK_prop}
\end{equation}
The Monte Carlo integration needed to evaluate the rhs of Eq. (\ref{eq:HHKK_prop})
is performed by generating a set of phase space points (\textbf{p}\textsubscript{0},\textbf{
q}\textsubscript{0}) according to the distribution function defined
by the overlap $\langle\mathbf{p}_{0},\mathbf{q}_{0}|\chi\rangle$.
$|{\bf p},{\bf q}\rangle$ indicates a coherent state, which has a
Gaussian representation in both position and momentum space. Specifically,

\begin{equation}
\left\langle \mathbf{x}|\mathbf{p},\mathbf{q}\right\rangle =\left(\frac{\text{det}(\pmb{\Gamma})}{\pi^{N_{vib}}}\right)^{1/4}e^{-\frac{1}{2}\left(\mathbf{x}-\mathbf{q}\right)^{T}\Gamma\left(\mathbf{x}-\mathbf{q}\right)+i\mathbf{p}^{T}\left(\mathbf{x}-\mathbf{q}\right)/\hbar},\label{eq:coherent_state}
\end{equation}
where $\pmb{\Gamma}$ is the \emph{N}\textsubscript{\emph{vib}}\emph{
\texttimes{} N}\textsubscript{\emph{vib}} coherent state width matrix
that is generally chosen to be diagonal with elements equal to the
harmonic frequencies of vibrations. The select phase space points
(\textbf{p}\textsubscript{0},\textbf{ q}\textsubscript{0}) serve
as starting conditions for classical dynamics runs that permit one
to evaluate the whole integrand. In fact, at every time \emph{t} along
the dynamics, the coherent state overlap $\langle\chi|\mathbf{p}_{t},\mathbf{q}_{t}\rangle$
is readily estimated; \emph{S}\textsubscript{\emph{t}}(\textbf{p}\textsubscript{0},\textbf{
q}\textsubscript{0}) is the instantaneous classical action, and \emph{C}\textsubscript{\emph{t}}(\textbf{p}\textsubscript{0},\textbf{
q}\textsubscript{0}) is the Herman-Kluk pre-exponential factor defined
as

\begin{equation}
C_{t}\left(\mathbf{p}_{0},\mathbf{q}_{0}\right)=\sqrt{\left|\frac{1}{2}\left(\frac{\partial\mathbf{q}_{t}}{\partial\mathbf{q}_{0}}+\pmb{\Gamma}^{-1}\frac{\partial\mathbf{p}_{t}}{\partial\mathbf{p}_{0}}\pmb{\Gamma}-i\hbar\frac{\partial\mathbf{q}_{t}}{\partial\mathbf{p}_{0}}\pmb{\Gamma}+\frac{i\pmb{\Gamma}^{-1}}{\hbar}\frac{\partial\mathbf{p}_{t}}{\partial\mathbf{q}_{0}}\right)\right|}.\label{eq:prefactor}
\end{equation}
The Fourier transform of Equation (\ref{eq:HHKK_prop}) yields the
SC power spectrum (i.e. the spectral density) associated to the vibrational
Hamiltonian $\hat{H}$ 

\begin{equation}
I(E)_{HK}=\dfrac{Re}{\pi\hbar}\int_{0}^{+\infty}dt\;e^{iEt/\hbar}\;\left\langle \chi\left|e^{-i\hat{H}t/\hbar}\right|\chi\right\rangle _{HK}.
\end{equation}
The SC estimates of the molecular eigenenergies of vibration correspond
to the energy values at which the peaks in the power spectrum are
localized. The ground state energy, i.e. the zero point energy, is
commonly shifted to zero and, in this case, the peaks in the power
spectrum are centered at the quantum (semiclassical) transition frequencies
of fundamentals and overtones. Unfortunately, convergence of the phase
space Monte Carlo integration for the Herman Kluk propagator is too
slow for application to sizeable molecular systems due to the oscillatory
nature of its integrand. This issue is overcome by Kaledin and Miller's
time average filter, which leads to a more efficient formula for the
spectral density known as the time averaged semiclassical initial
value representation (TA SCIVR)

\begin{equation}
I(E)_{TASCIVR}=\left(\frac{1}{2\pi\hbar}\right)^{N_{vib}}\iintop d\mathbf{p}_{0}d\mathbf{q}_{0}\frac{1}{2\pi\hbar T}\left|\intop_{0}^{T}e^{\frac{i}{\hbar}\left[S_{t}\left(\mathbf{p}_{0},\mathbf{q}_{0}\right)+Et+\phi_{t}\left(\mathbf{p}_{0},\mathbf{q}_{0}\right)\right]}\langle\chi|\mathbf{p}_{t},\mathbf{q}_{t}\rangle dt\right|^{2}.\label{eq:TASCIVR}
\end{equation}
In Eq.(\ref{eq:TASCIVR}) $\phi_{t}\left(\mathbf{p}_{0},\mathbf{q}_{0}\right)$
is the phase of the original complex-valued Herman-Kluk pre-exponential
factor introduced owing to the so-called ``separable'' approximation,\citep{Kaledin_Miller_Timeaveraging_2003}
and \emph{T} is the total time of the dynamics. The key advantage
of Eq. (\ref{eq:TASCIVR}) is that the integrand is positive definite
and this facilitates the numerical convergence of the integration.
The derivation of Eq. (\ref{eq:TASCIVR}) is discussed extensively
in Ref. \citenum{Kaledin_Miller_Timeaveraging_2003}, and the interested
reader can find all the details therein. 

For the purposes of this work it is important to point out that Eq.
(\ref{eq:TASCIVR}) is the basic working formula for our spectral
simulations. It has been applied previously to several molecules and
further developed to interface the TA-SCIVR formalism to ``on-the-fly''
dynamics for application to much larger molecular systems. Specifically,
on-the-fly simulations must rely on a limited number of trajectories
to be computationally affordable. This is achieved by adopting for
each vibrational state a few tailored trajectories (i.e. SC propagators)
and reference states $|\chi\rangle$, a technique known as multiple
coherent semiclassical initial value representation (MC SCIVR). The
trajectories are run at an energy corresponding to the harmonic estimate
for the generic \emph{k}-th state to investigate, while the corresponding
reference state is chosen as 

\begin{equation}
\left|\chi^{(k)}\right\rangle =\prod_{i=1}^{N_{vib}}\sum_{j=1}^{N_{\alpha}}\pmb{\varepsilon}_{ij}^{(k)}\left|p_{eq,i}^{(k)},q_{eq,i}^{(k)}\right\rangle .\label{eq:MCTASCIVR}
\end{equation}
The coherent states in Eq.(\ref{eq:MCTASCIVR}) are centered at specific
positions and momenta $({\bf p}_{eq}^{(k)},{\bf q}_{eq}^{(k)})$,
which determine the initial conditions and the energy of the classical
trajectories needed by the semiclassical simulation. \emph{$\text{\text{N}}_{\alpha}$}
accounts for the possible duplication of coherent states, which are
linearly combined through the array of integers ($\pmb{\varepsilon}$),
to enforce parity and molecular symmetries. Parity symmetry allows
to enhance specific peaks in the power spectrum (i.e. the ground state,
specific fundamental transitions, or overtones) and it is obtained
by inverting the sign of the momentum in the duplicate coherent state.
Molecular symmetry (detection of peaks corresponding to vibrations
belonging to a specific irreducible representation of the symmetry
group of the molecule) is introduced by inverting the sign of the
(equilibrium) positions in the duplicate coherent state.\citep{Kaledin_Miller_TAmolecules_2003,Ceotto_AspuruGuzik_Multiplecoherent_2009}
The greatest novelty of the MC SCIVR procedure is that accurate results
can be obtained by running just one tailored trajectory per state
with the effect of alleviating the computational cost substantially. 

Another problem related to Eq. (\ref{eq:TASCIVR}) is that the coherent
state overlap $\langle\chi|\mathbf{p}_{t},\mathbf{q}_{t}\rangle$
is more and more likely to give a negligible contribution as \emph{N}\textsubscript{\emph{vib}}
increases, so that a spectrally resolved Fourier signal is harder
and harder to collect. A very recent advance able to overcome this
problem is the divide-and-conquer semiclassical initial value representation
(DC SCIVR), which is based on performing semiclassical calculations
in lower dimensional subspaces. The dynamics is still performed in
full dimensionality permitting one to recover at least partially the
interaction between different subspaces, and Eq. (\ref{eq:TASCIVR})
remains the working formula. The several quantities appearing in the
integrand of Eq. (\ref{eq:TASCIVR}) are readily projected onto the
subspaces with the exception of the action, which requires a redefinition
of the projected potential energy due to its non-separability in general.
For an \emph{M}-dimensional subspace 

\begin{equation}
\tilde{I}(E)_{DCSCIVR}=\left(\frac{1}{2\pi\hbar}\right)^{M}\iintop d\tilde{\mathbf{p}}_{0}d\tilde{\mathbf{q}}_{0}\frac{1}{2\pi\hbar T}\left|\intop_{0}^{T}e^{\frac{i}{\hbar}\left[\tilde{S}_{t}\left(\tilde{\mathbf{p}}_{0},\tilde{\mathbf{q}}_{0}\right)+Et+\tilde{\phi}_{t}\left(\tilde{\mathbf{p}}_{0},\tilde{\mathbf{q}}_{0}\right)\right]}\langle\tilde{\chi}|\tilde{\mathbf{p}}_{t},\tilde{\mathbf{q}}_{t}\rangle dt\right|^{2}.\label{eq:DCSCIVR}
\end{equation}

\begin{eqnarray}
 &  & \tilde{S}_{t}(\tilde{\mathbf{p}}_{0},\tilde{\mathbf{q}}_{0})=\int_{0}^{t}dt^{\prime}\,(\tilde{T}(t^{\prime})-\tilde{V}(t^{\prime})),\nonumber \\
 &  & \tilde{V}(t^{\prime})=V(\tilde{q}_{M}(t^{\prime});q_{N_{vib}-M}^{eq})+\lambda(t^{\prime});\nonumber \\
 &  & \lambda(t)=V(\tilde{q}_{M}(t);q_{N_{vib}-M}(t))-[V(\tilde{q}_{M}^{eq};q_{N_{vib}-M}(t)+V(\tilde{q}_{M}(t);q_{N_{vib}-M}^{eq})];\nonumber \\
 &  & \tilde{V}(t^{\prime})=V(\tilde{q}_{M}(t^{\prime});q_{N_{vib}-M}(t^{\prime}))-V(\tilde{q}_{M}^{eq};q_{N_{vib}-M}(t^{\prime})).\label{eq:Proj_Action}
\end{eqnarray}

Projected quantities in Eqs. (\ref{eq:DCSCIVR}) and (\ref{eq:Proj_Action})
are indicated by tildes. Upon parametrization of the degrees of freedom
external to the \emph{M}-dimensional subspace and correction by means
of the external time-dependent field $\lambda(t),$ the projected
potential is obtained as a difference between the instantaneous potential
and the potential at the corresponding unrelaxed subspace minimum.
Combination of the multiple coherent and divide-and-conquer techniques,
named MC-DC SCIVR is feasible and effective. A complete treatment
of DC SCIVR can be found in Refs.\citenum{ceotto_conte_DCSCIVR_2017, DiLiberto_Ceotto_Jacobiano_2018},
where appropriate strategies for determining the best suited subspaces
have been illustrated and discussed. 

A final issue concerning Eq. (\ref{eq:TASCIVR}) is that evolution
of the phase $\phi_{t}\left(\mathbf{p}_{0},\mathbf{q}_{0}\right)$,
which bears key quantum effects, requires to calculate the Hessian
matrix along the whole classical trajectory and the computational
effort therefore increases rapidly with the number of degrees of freedom.
This is the bottleneck of semiclassical calculations plaguing both
full dimensional and DC-SCIVR approaches. To overcome the problem
it is necessary to find a way to reduce the number of Hessian calculations
while keeping a sufficient accuracy for vibrational frequencies. A
first effort in this direction has been made by developing a compact
finite-difference approximation.\citep{Ceotto_Hase_AcceleratedSC_2013}
If the Hessian is calculated at the \emph{i}-th step of the dynamics
($\mathcal{H}_{i}$), then the approximation estimates $\mathcal{H}_{j}$
(\emph{j}=\emph{i}+1,...,\emph{i+N}-1) at the following \emph{N}-1
steps according to the formula

\begin{equation}
\mathcal{\mathcal{H}}_{j}=\mathcal{H}_{j-1}+(1-\lambda)\dfrac{\boldsymbol{R}\boldsymbol{R}^{T}}{\boldsymbol{R}^{T}\Delta\boldsymbol{q}}+\lambda\left(\dfrac{\Delta\boldsymbol{q}\boldsymbol{R}^{T}+\boldsymbol{R}^{T}\Delta\boldsymbol{q}}{||\Delta\boldsymbol{q}||^{2}}-\dfrac{\boldsymbol{R}^{T}\Delta\boldsymbol{q}}{||\Delta\boldsymbol{q}||^{4}}\Delta\boldsymbol{q}\Delta\boldsymbol{q}^{T}\right),\label{eq:CFD}
\end{equation}
where $\lambda$ is a parameter identifying different families of
approximations according to its value; $\Delta$\textbf{q} = \textbf{q}\textsubscript{j}
- \textbf{q}\textsubscript{i}, and \textbf{R} = 2(\textbf{g}\textsubscript{j}
- \textbf{g}\textsubscript{i} - $\mathcal{H}$\textsubscript{i}$\Delta$\textbf{q)}
with \textbf{g}\textsubscript{i} being the gradient at the \emph{i}-th
geometry. The simplest form of the approximation employs the value
$\lambda$ = 1, and it is known as the Hessian Update (HU) Powell
Symmetric Broyden scheme. A more refined choice of $\lambda$ leads
to the HU Bofill approximation. Calculation of the Hessian matrix
is performed only at intervals of \emph{N} steps, while, according
to Eq. (\ref{eq:CFD}), only the computationally cheaper gradient
must be evaluated for all the other steps. These schemes have been
shown to work well for small-to-medium size molecules.\citep{Wu_Zhuang_HessApprox_2010,Zhuang_Ceotto_Hessianapprox_2012,Ceotto_Hase_AcceleratedSC_2013} 

In this work we introduce an alternative strategy to decrease the
number of Hessian calculations. It is based on the dynamical creation
of a database of Hessians (DBH) and related geometries, which can
be exploited whenever the instantaneous molecular geometry is close
enough to one of the $N_{db}$ records already present in the database.
An estimate of the ``distance'' between two molecular configurations
can be obtained either by evaluating the root mean square difference
between vibrational normal mode coordinates in the two configurations,

\begin{equation}
\sigma\,({\bf q},{\bf q}_{db})=\sqrt{\dfrac{\sum_{l=1}^{N_{vib}}(q_{l}-q_{l,db})^{2}}{N_{vib}}},\label{eq:RMS}
\end{equation}
where $q_{l}$ and $q_{l,db}$ are the values of the \emph{l}-th normal
mode for the instantaneous geometry and one included in the database
respectively, or by estimating the absolute difference for each mode 

\begin{equation}
\Theta{}_{l}\,({\bf q},{\bf q}_{db})=|q_{l}-q_{l,db}|,\quad l=1,\ldots,N_{vib}.\label{eq:sigma}
\end{equation}
To determine whether the two configurations are enough close or not,
an arbitrary threshold $\rho$ is introduced. It is straightforward
to demonstrate that $\sigma\,({\bf q},{\bf q}_{db})<\rho$ is a weaker
condition than having the $\Theta\,({\bf q},{\bf q}_{db})$ test passed
for each mode, i.e. $\Theta_{l}\,({\bf q},{\bf q}_{db})<\rho\quad\forall l\in[1,N_{db}]$.
For this reason we employ Eq. (\ref{eq:sigma}) to determine the set
of suitable records in the database, while Eq.(\ref{eq:RMS}) is used
to identify the closest among database candidates (${\bf q}_{db}^{closest}$)
to the instantaneous geometry. 

The search for the best match among the \emph{N}\textsubscript{\emph{db}}
records ${\bf q}$\textsubscript{\emph{db}} in the database for the
instantaneous geometry ${\bf q}$ is another important aspect of the
method since it may become computationally costly when the size of
the database grows large. To accelerate the search, mode-1 components
of geometries in the database are sorted in increasing order. The
algorithm starts with calculation of $\Theta_{1}\,({\bf q},{\bf q}_{db})$
along the database and check of the $\Theta_{1}\,({\bf q},{\bf q}_{db})<\rho$
condition. Once the test is passed, the search based on the first
normal mode is stopped as soon as $\Theta_{1}\,({\bf q},{\bf q}_{db})>\rho$.
Then, the $\Theta_{l}\,({\bf q},{\bf q}_{db})<\rho$ condition is
evaluated sequentially for the other normal modes only on the restricted
set of records that have passed all previous $\Theta_{l^{\prime}}\,({\bf q},{\bf q}_{db})<\rho,\,l^{\prime}=1,\ldots,l-1$
tests. 

\begin{figure}[b]
\begin{centering}
\includegraphics[scale=0.4]{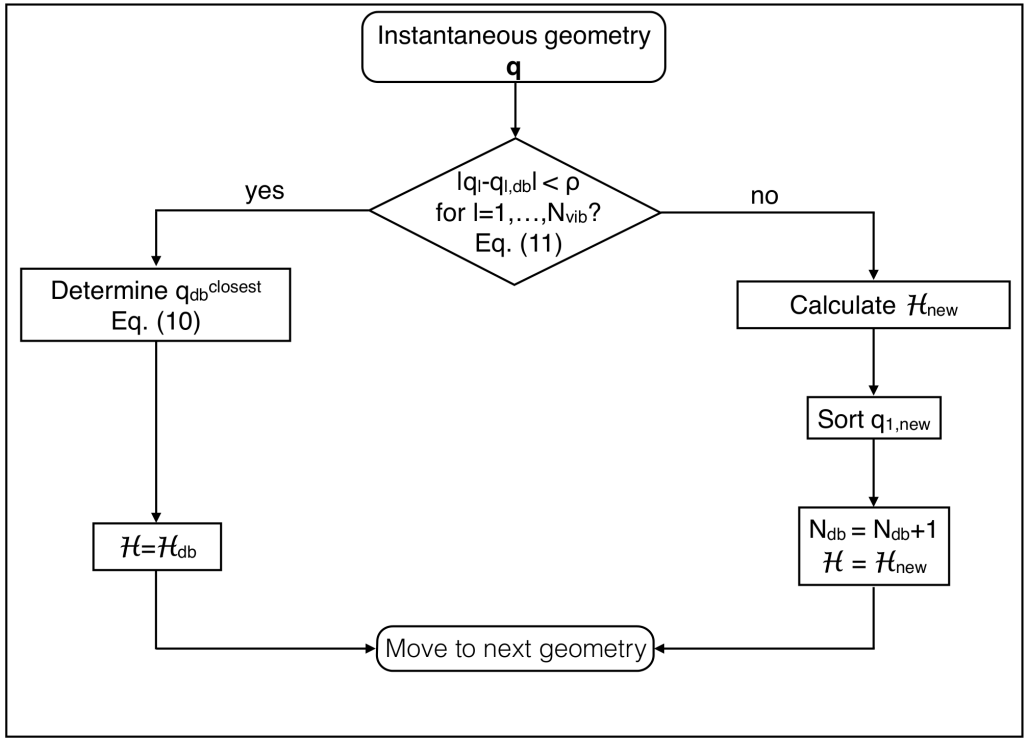}
\par\end{centering}
\caption{Flow chart for the DBH method.\label{Flowchart_DBH}}
\end{figure}

If, after all \emph{N}\textsubscript{\emph{vib}} modes have been
examined, there are one or more configurations still left, then the
Hessian associated with the geometry with the smallest $\sigma\,({\bf q},{\bf q}_{db})$
value, i.e. ${\bf q}_{db}^{closest}$, is selected to approximate
the instantaneous Hessian in the SC simulation. Otherwise, a new Hessian
calculation is performed and added to the database in the correct
\emph{k}-th position as determined by its\textbf{ ${\bf q}_{1,db}$}
value. Figure \ref{Flowchart_DBH} shows the flow chart for the method.

 Some preliminary tests on systems of increasing dimensionality have
been undertaken to check on the performance of the approach depending
on database size, threshold $\rho$, and system complexity. As a first
test, we employed analytical water\citep{Bowman_Zuniga_H2Opotential_1988}
and methane\citep{lee_taylor_PESch4_1995} potentials to construct
three databases of different sizes by running different numbers of
trajectories. Then, an additional trajectory was run to evaluate,
at a specific value of $\rho$: i) the number of geometries for which
a suitable match in the database was found; ii) the root mean square
deviation ($\sigma_{Hess}$) of the elements of the associated Hessian
in the database from those calculated at the trajectory instantaneous
geometry. $\sigma$\textsubscript{Hess} is defined as 

\begin{equation}
\sigma_{Hess}=\sqrt{\dfrac{\sum_{m}^{N_{vib}}\sum_{n}^{N_{vib}}(\mathcal{\mathcal{H}}_{mn}-\mathcal{\mathcal{H}}_{mn,db})^{2}}{N_{vib}-1}},\label{RMS_Hess}
\end{equation}
where $\mathcal{\mathcal{H}}_{mn}$ is the (\emph{m,n}) component
of the Hessian matrix. A single test trajectory is representative
of the general case of interest for SC simulations in which initial
conditions are sampled around the equilibrium geometry.

Table \ref{tab:H2O_test} reports the results. Each trajectory was
evolved in full dimensionality with a symplectic algorithm for a total
of 2,500 steps, i.e. it is made of 2,500 geometries. The larger the
database becomes, the easier it is to find a suitable record in it
and the more accurate the Hessian database estimate.

\begin{table}[H]
\caption{Comparison of different databases for H\protect\textsubscript{2}O
and CH\protect\textsubscript{4}. The threshold parameter is $\rho$
= 1 for H\protect\textsubscript{2}O and $\rho$ = 10 for CH\protect\textsubscript{4}.
\emph{N}\protect\textsuperscript{\emph{db}} indicates the number
of geometries constituting the database; \emph{N}\protect\textsuperscript{\emph{mg}}
is the number of geometries of the test trajectory for which a matching
database entry is found; $\sigma_{Hess}$ is the root mean square
deviation of the Hessian elements for the matched geometries, as defined
in Eq. (\ref{RMS_Hess}).\label{tab:H2O_test}}

\renewcommand{\arraystretch}{1.2}
\centering{}%
\begin{tabular}{|c|c|c|c|c|c|}
\hline 
\multicolumn{3}{|c|}{H\textsubscript{2}O} & \multicolumn{3}{c|}{CH\textsubscript{4}}\tabularnewline
\hline 
N\textsuperscript{db} & N\textsuperscript{mg} & $\sigma_{Hess}$ & N\textsuperscript{db} & N\textsuperscript{mg} & $\sigma_{Hess}$\tabularnewline
\hline 
\hline 
2500 & 295 & 9.4$\,$10\textsuperscript{-6} & 2500 & 118 & 5.3\,10\textsuperscript{-5}\tabularnewline
\hline 
25000 & 2157 & 8.5$\,$10\textsuperscript{-6} & 25000 & 1209 & 4.1\,10\textsuperscript{-5}\tabularnewline
\hline 
250000 & 2500 & 3.0$\,$10\textsuperscript{-6} & 250000 & 2500 & 2.6\,10\textsuperscript{-5}\tabularnewline
\hline 
\end{tabular}
\end{table}
For given H\textsubscript{2}O and CH\textsubscript{4} databases,
constructed by running ten trajectories made of 2,500 steps each (i.e.
a database made of 25,000 records), the same quantities have been
estimated for different values of the threshold parameter. The results
are shown in Table \ref{tab:H2O_test_2}. It is not surprising to
find that, for a given database, a stricter threshold value leads
to fewer matched geometries and also to enhanced accuracy. 

\begin{table}[H]
\caption{Number of matched geometries (\emph{N}\protect\textsuperscript{\emph{mg}})
and root mean square Hessian deviation $\sigma_{Hess}$ at various
threshold values for a 25000-point database for H\protect\textsubscript{2}O
and CH\protect\textsubscript{4}. \label{tab:H2O_test_2}}

\renewcommand{\arraystretch}{1.2}
\centering{}%
\begin{tabular}{|c|c|c|c|c|c|}
\hline 
\multicolumn{3}{|c|}{H\textsubscript{2}O} & \multicolumn{3}{c|}{CH\textsubscript{4}}\tabularnewline
\hline 
$\rho$ & N\textsuperscript{mg} & $\sigma_{Hess}$ & $\rho$ & N\textsuperscript{mg} & $\sigma_{Hess}$\tabularnewline
\hline 
\hline 
10 & 2500 & 9.4$\,$10\textsuperscript{-6} & 50 & 2500 & 4.4\,10\textsuperscript{-5}\tabularnewline
\hline 
1 & 2157 & 8.5$\,$10\textsuperscript{-6} & 10 & 1209 & 4.1\,10\textsuperscript{-5}\tabularnewline
\hline 
0.1 & 17 & 1.2$\,$10\textsuperscript{-6} & 5 & 45 & 2.3\,10\textsuperscript{-5}\tabularnewline
\hline 
\end{tabular}
\end{table}
Similar trends have been obtained in an application to glycine, which
requires ``on-the-fly'' (i.e. direct) dynamics. The database had
to be constructed from a single trajectory due to the computational
overhead of direct dynamics. Specifically, given a 5000-step trajectory
started with harmonic ZPE, we employed the first 2500 steps to build
the database and the final 2500 steps as our test trajectory. The
trajectory was calculated at DFT/B3LYP level of theory with an aug-cc-pVDZ
basis set. An interesting feature in an ``on-the-fly'' application
is that the cost of the search algorithm versus the time needed to
run a single trajectory and build a database with \textit{ab initio}
calculated Hessians is negligible. This is different from what happens
for smaller molecules for which a fast-to-evaluate PES is available.
The same conclusion is valid and reinforced when applying DBH to the
much larger Ac-Phe-Met-NH$_{2}$ molecule at DFT-B3LYP-D level of
theory and 6-31G{*} basis set. Table \ref{tab:time_ratio} reports
some insightful data.

\begin{table}[H]
\caption{Comparison between the time needed to run a first trajectory (single
core, 2500 steps) with creation of the associate ordered database
(t\protect\textsubscript{traj+db}) and the time needed to search
(t\protect\textsubscript{search}) in the database during a second
test trajectory. Times are reported for molecules of increasing dimensionality.\label{tab:time_ratio}}

\renewcommand{\arraystretch}{1.1}
\centering{}%
\begin{tabular}{|c|c|c|c|c|}
\hline 
Molecule & N\textsubscript{vib} & t\textsubscript{traj+db}(s) & t\textsubscript{search}(s) & t\textsubscript{search}/t\textsubscript{traj+db}\tabularnewline
\hline 
\hline 
H\textsubscript{2}O & 3 & 0.984 & 0.5 & 0.508\tabularnewline
\hline 
CH\textsubscript{4}\textsuperscript{} & 9 & 4.663 & 1.5 & 0.322\tabularnewline
\hline 
Glycine & 24 & > 10\textsuperscript{5} & < 10 & $\approx$ 0\tabularnewline
\hline 
Ac-Phe-Met-NH$_{2}$ & 132 & > 10\textsuperscript{6} & < 60 & $\approx$ 0\tabularnewline
\hline 
\end{tabular}
\end{table}

The way the Hessian database is built and employed depends on the
type of dynamics adopted. When an analytical PES is available, many
trajectories can be run to integrate Eq. (\ref{eq:TASCIVR}) and each
step of the dynamics is evolved by means of a 4-step symplectic algorithm.\citep{Brewer_Manolopoulos_15dof_1997}
The search for a suitable Hessian in the database (or the calculation
of a new Hessian, according to the flow chart in Fig. \ref{Flowchart_DBH})
is performed at the first of the four symplectic steps. For the three
remaining steps, the HU Bofill scheme is invoked. Clearly, HU Bofill
can be used along the dynamics to restrict the search in the database
to just a small fraction of steps. Therefore, we adopt the expression
HU=N to indicate that the search is performed every N dynamics steps.

In the case of an \textit{ab initio} on-the-fly simulation the symplectic
velocity-Verlet integrator is used for the dynamics and Hessian calculations
are undertaken only once the entire dynamics is complete. This allows
the construction of a ``predictor'' code that, given the dynamics
and a chosen threshold $\rho$, and following the flow chart of Fig.
\ref{Flowchart_DBH}, returns the geometries at which the Hessian
must be calculated \emph{ab initio}. For all other geometries, a suitable
match in the database is available ($\mathcal{H}=\mathcal{H}_{db}$).

\section{Results\label{sec:Results}}

\subsection{Methane}

First we present an application of the Hessian database approach to
the small methane molecule, for which an analytical potential energy
surface is available.\citep{lee_taylor_PESch4_1995} We performed
several simulations, in which DBH was employed either at each dynamics
step (DBH; HU=1) or every N dynamics steps when interfaced to Hessian
Update schemes (DBH; HU=N, N>1). As anticipated in Section \ref{sec:Theory},
if no suitable database record was available, then a new Hessian matrix
was calculated by finite differences at the first of the four symplectic
steps and added to the database. For comparison purposes, additional
simulations were also undertaken without using DBH (no DBH). Each
simulation consisted of 33000 trajectories evolved for 2500 steps
with a time step of 10 a.u. (for a total of about 600 fs). Chaotic
trajectories were discarded before the end of the propagation on the
basis of the value of the determinant of the monodromy matrix $M$.
Specifically, a trajectory was rejected whenever the deviation of
the determinant of $M^{T}M$ from its expected value of unity\citep{Kaledin_Miller_TAmolecules_2003}
had become greater than 1\%. The different databases were constructed
from the first 2000 trajectories of each simulation employing the
search threshold $\rho=5$, which led to database sizes of several
thousand records as reported in Table \ref{tab:Percentage-of-trajectory}.
\begin{table}[H]
\caption{Application of DBH to methane with threshold $\rho=5$ and different
periodicities N of the Hessian Update Bofill scheme (HU=N). The size
of the database is reported in the second column (DB Size); column
3 shows the percentage of trajectory rejection (\% Rejection) given
a det($M^{T}M$) tolerance of 1\%; in the last column, the percentage
of rejection is reported for a simulation that does not employ DBH.\label{tab:Percentage-of-trajectory}}

\renewcommand{\arraystretch}{1.2}
\centering{}%
\begin{tabular}{|c|c|c|c|}
\hline 
 & \multicolumn{2}{c|}{DBH} & No DBH\tabularnewline
\hline 
\hline 
 & DB Size & \% Rejection & \% Rejection\tabularnewline
\hline 
HU=1 & 59277 & 58.3 & 54.7\tabularnewline
\hline 
HU=2 & 55082 & 63.7 & 65.6\tabularnewline
\hline 
HU=5 & 24997 & 79.6 & 80.8\tabularnewline
\hline 
HU=10 & 9899 & 94.4 & 94.8\tabularnewline
\hline 
\end{tabular}
\end{table}
\noindent The Table lists the percentage of rejected trajectories
for simulations involving the HU Bofill scheme in the presence and
absence of the DBH procedure. The percentages are similar in the two
instances. 

\begin{figure}[H]
\begin{centering}
\includegraphics[scale=0.62]{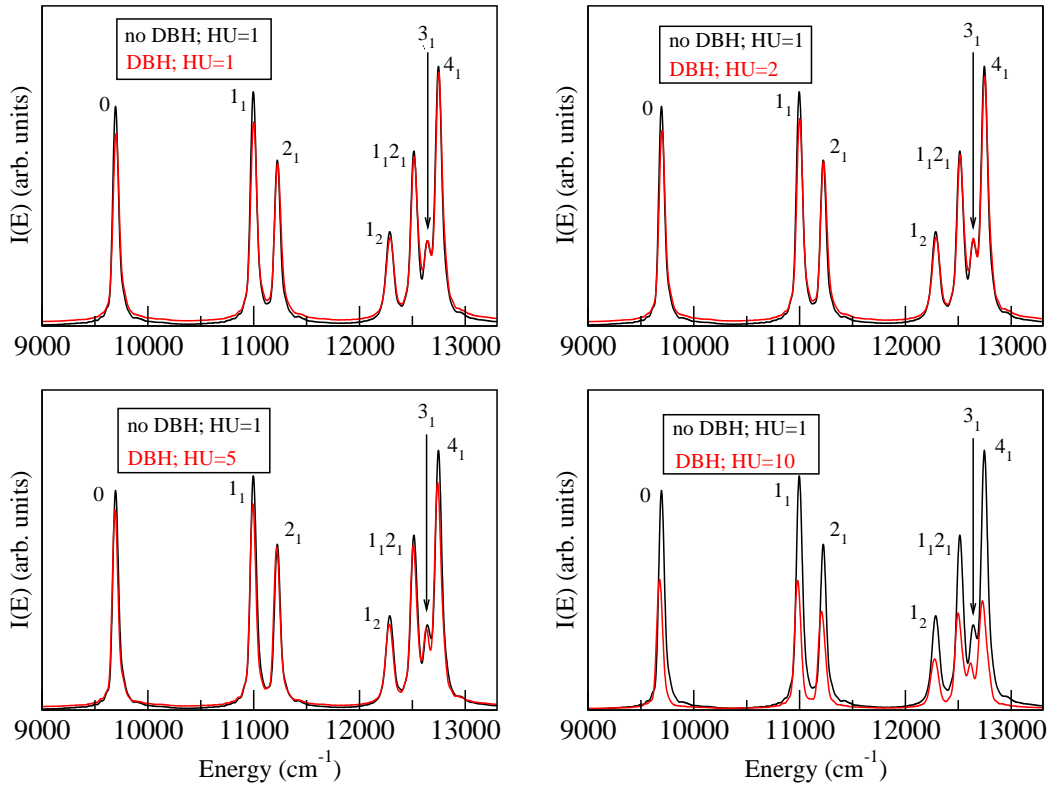}
\par\end{centering}
\caption{Comparison between the power spectrum obtained for CH\protect\textsubscript{4}
calculating all Hessians (no DBH; HU=1, black line) and those employing
the DBH technique (top left panel, red line) and DBH in connection
with Hessian Update schemes with different periodicities (other panels,
red lines).\label{fig_CH4}}
\end{figure}

Figure \ref{fig_CH4} shows a comparison between a simulation in which
the Hessian has been calculated at all dynamics steps (no DBH; HU=1)
and the four DBH simulations sketched in Table \ref{tab:Percentage-of-trajectory}.
In all these simulations, a Husimi distribution of initial phase-space
conditions around the harmonic zero point energy was employed, while
a coherent state centered at the equilibrium geometry and at the momentum
corresponding to the harmonic ZPE was chosen as a reference state.

An examination of the frequencies of vibration reported in Table \ref{tab:Vibrational-frequencies-CH4}
leads to the observation that results obtained by means of the database
technique are in good agreement with the standard approach in which
all Hessians are calculated. The accuracy starts to diminish, especially
regarding the ZPE value, for the DBH; HU=10 case. This was already
clear from the bottom right panel of Figure \ref{fig_CH4}, in which
a loss in signal intensity is detected, an aspect related to the high
percentage of rejected trajectories. The full widths at half maximum
(FWHM) of the peaks are reported to provide a rough estimate of data
uncertainty, and frequency estimates are generally well within the
tolerance interval. 

\begin{table}[H]
\caption{Vibrational frequencies and ZPE for CH\protect\textsubscript{4} in
cm\protect\textsuperscript{-1}. FWHM values (cm\protect\textsuperscript{-1})
are given in parentheses. Column 1 indicates the vibrational transition
(0 represents the ground state). Column 2 is reserved to the quantum
mechanical (QM) benchmark values from Ref.\citenum{Carter_Bowman_Methane_1999}.
Column 3 reports the outcomes of the all-Hessian simulation. The last
four columns contain the results of DBH simulations. In the last row
the mean absolute error (MAE) in cm\protect\textsuperscript{-1} with
respect to QM results is given. FWHM values are not available for
the 0$\rightarrow$ 3\protect\textsubscript{1} transition.\label{tab:Vibrational-frequencies-CH4}}

\renewcommand{\arraystretch}{1.4}
\centering{}%
\begin{tabular}{|c|c|c|c|c|c|c|}
\hline 
Transition & QM\textsuperscript{} & no DBH; HU=1 & DBH; HU=1 & DBH; HU=2 & DBH; HU=5 & DBH; HU=10\tabularnewline
\hline 
\hline 
0$\rightarrow$ 1\textsubscript{1} & 1313 & 1302 (68) & 1301 (72)  & 1301 (73) & 1303 (69) & 1304 (71)\tabularnewline
\hline 
0$\rightarrow$ 2\textsubscript{1} & 1535 & 1529 (69) & 1528 (77) & 1528 (75) & 1529 (71) & 1529 (80)\tabularnewline
\hline 
0$\rightarrow$ 1\textsubscript{2} & 2624 & 2592 (87) & 2590 (95) & 2591 (92) & 2592 (92) & 2599 (91)\tabularnewline
\hline 
0$\rightarrow$ 1\textsubscript{1}2\textsubscript{1} & 2836 & 2820 (78) & 2818 (83) & 2818 (80) & 2819 (78) & 2820 (85)\tabularnewline
\hline 
0$\rightarrow$ 3\textsubscript{1} & 2949 & 2948 & 2945 & 2943 & 2942 & 2938\tabularnewline
\hline 
0$\rightarrow$ 4\textsubscript{1} & 3053 & 3053 (76) & 3051 (79) & 3050 (80) & 3050 (81) & 3050 (93)\tabularnewline
\hline 
ZPE & 9707 & 9694 (65) & 9699 (71) & 9699 (69) & 9690 (65) & 9677 (68)\tabularnewline
\hline 
MAE & - & 11.3 & 12.1 & 12.4 & 13.1 & 14.3\tabularnewline
\hline 
\end{tabular}
\end{table}

\subsection{Glycine}

Moving to ``on-the-fly'' applications, we employed the database
approach to study the vibrational frequencies of glycine in full dimensionality
(24 degrees of freedom). Calculations were based on a single, 2500-step
long trajectory with a time step of 10 a.u. \emph{Ab initio} molecular
dynamics was performed at harmonic ZPE at DFT-B3LYP level of theory
with aVDZ basis set, in agreement with a previous semiclassical study
of glycine.\citep{Gabas_Ceotto_Glycine_2017} For the test of the
determinant of the monodromy matrix we used a threshold equal to 10\textsuperscript{-2}.
Furthermore, a well-established regularization technique for the monodromy
matrix\citep{DiLiberto_Ceotto_Prefactors_2016} was adopted to avoid
discard of the trajectory before its scheduled end. As anticipated
in Section \ref{sec:Theory}, when performing SC spectroscopy ``on-the-fly'',
Hessian matrices are calculated once the entire trajectory has already
been determined. This allows one to exploit DBH to reduce the number
of Hessian calculations similarly to HU schemes but in a more flexible
way. 

\begin{table}[H]
\caption{{\small{}ZPE energies and frequencies of fundamental transitions of
glycine (cm}\protect\textsuperscript{{\small{}-1}}{\small{}). MAE
values (cm}\protect\textsuperscript{{\small{}-1}}{\small{}) are
determined with respect to the SC simulation with all Hessians calculated
}\emph{\small{}ab initio}{\small{} (all Hess). Columns 3-5 report
the outcomes of DBH calculations based on several sets of }\emph{\small{}ab
initio}{\small{} Hessians. The last three columns provide estimates
obtained with the HU scheme. Numbers in parentheses are the FWHM (cm}\protect\textsuperscript{{\small{}-1}}{\small{})
of the associate Fourier transform signal and their average is reported
in the last row.\label{tab-DB-Gly}}}

\renewcommand{\arraystretch}{1.4}
\centering{}%
\begin{tabular}{|c|c|c|c|c|c|c|c|}
\hline 
{\footnotesize{}Mode} & {\footnotesize{}All Hess} & {\footnotesize{}DBH = 3.89} & {\footnotesize{}DBH = 7.62} & {\footnotesize{}DBH = 12.69 } & {\footnotesize{}HU = 4} & {\footnotesize{}HU = 8 } & {\footnotesize{}HU=13}\tabularnewline
\hline 
\hline 
{\footnotesize{}24} & {\footnotesize{}3639 (55)} & {\footnotesize{}3639 (56)} & {\footnotesize{}3639 (60)} & {\footnotesize{}3639 (59)} & {\footnotesize{}3639 (55)} & {\footnotesize{}3636 (60)} & {\footnotesize{}3637 (73)}\tabularnewline
\hline 
{\footnotesize{}23} & {\footnotesize{}3372 (59)} & {\footnotesize{}3372 (83)} & {\footnotesize{}3369 (60)} & {\footnotesize{}3369 (48)} & {\footnotesize{}3372 (60)} & {\footnotesize{}3363 (70)} & {\footnotesize{}3370 (88)}\tabularnewline
\hline 
{\footnotesize{}22} & {\footnotesize{}3375 (58)} & {\footnotesize{}3378 (59)} & {\footnotesize{}3375 (61)} & {\footnotesize{}3378 (67)} & {\footnotesize{}3375 (59)} & {\footnotesize{}3372 (64)} & {\footnotesize{}3370 (82)}\tabularnewline
\hline 
{\footnotesize{}21} & {\footnotesize{}2904 (59)} & {\footnotesize{}2904 (59)} & {\footnotesize{}2904 (61)} & {\footnotesize{}2904 (62)} & {\footnotesize{}2901 (59)} & {\footnotesize{}2898 (61)} & {\footnotesize{}2895 (72)}\tabularnewline
\hline 
{\footnotesize{}20} & {\footnotesize{}2904 (53)} & {\footnotesize{}2907 (53)} & {\footnotesize{}2901 (55)} & {\footnotesize{}2904 (57)} & {\footnotesize{}2904 (54)} & {\footnotesize{}2901 (57)} & {\footnotesize{}2902 (78)}\tabularnewline
\hline 
{\footnotesize{}19} & {\footnotesize{}1779 (54)} & {\footnotesize{}1779 (53)} & {\footnotesize{}1782 (61)} & {\footnotesize{}1779 (58)} & {\footnotesize{}1779 (54)} & {\footnotesize{}1779 (60)} & {\footnotesize{}1777 (60)}\tabularnewline
\hline 
{\footnotesize{}18} & {\footnotesize{}1662 (52)} & {\footnotesize{}1662 (52)} & {\footnotesize{}1662 (46)} & {\footnotesize{}1668 (47)} & {\footnotesize{}1662 (50)} & {\footnotesize{}1656 (51)} & {\footnotesize{}1645 (60)}\tabularnewline
\hline 
{\footnotesize{}17} & {\footnotesize{}1404 (52)} & {\footnotesize{}1404 (53)} & {\footnotesize{}1404 (66)} & {\footnotesize{}1401 (57)} & {\footnotesize{}1404 (54)} & {\footnotesize{}1401 (63)} & {\footnotesize{}1405 (61)}\tabularnewline
\hline 
{\footnotesize{}16} & {\footnotesize{}1380 (66)} & {\footnotesize{}1383 (69)} & {\footnotesize{}1386 (68)} & {\footnotesize{}1383 (78)} & {\footnotesize{}1380 (67)} & {\footnotesize{}1383 (66)} & {\footnotesize{}1378 (62)}\tabularnewline
\hline 
{\footnotesize{}15} & {\footnotesize{}1344 (55)} & {\footnotesize{}1347 (56)} & {\footnotesize{}1350 (59)} & {\footnotesize{}1347 (60)} & {\footnotesize{}1344 (56)} & {\footnotesize{}1347 (56)} & {\footnotesize{}1342 (57)}\tabularnewline
\hline 
{\footnotesize{}14} & {\footnotesize{}1287 (85)} & {\footnotesize{}1287 (95)} & {\footnotesize{}1281 (68)} & {\footnotesize{}1287 (100)} & {\footnotesize{}1284 (86)} & {\footnotesize{}1275 (75)} & {\footnotesize{}1258 (78)}\tabularnewline
\hline 
{\footnotesize{}13} & {\footnotesize{}1158 (56)} & {\footnotesize{}1161 (57)} & {\footnotesize{}1164 (67)} & {\footnotesize{}1158 (64)} & {\footnotesize{}1158 (57)} & {\footnotesize{}1161 (60)} & {\footnotesize{}1159 (60)}\tabularnewline
\hline 
{\footnotesize{}12} & {\footnotesize{}1122 (53)} & {\footnotesize{}1125 (53)} & {\footnotesize{}1131 (57)} & {\footnotesize{}1134 (43)} & {\footnotesize{}1122 (53)} & {\footnotesize{}1125 (53)} & {\footnotesize{}1120 (51)}\tabularnewline
\hline 
{\footnotesize{}11} & {\footnotesize{}1098 (59) } & {\footnotesize{}1101 (60)} & {\footnotesize{}1101 (62)} & {\footnotesize{}1098 (68)} & {\footnotesize{}1098 (59)} & {\footnotesize{}1098 (63)} & {\footnotesize{}1096 (71)}\tabularnewline
\hline 
{\footnotesize{}10} & {\footnotesize{}900 (55)} & {\footnotesize{}900 (55)} & {\footnotesize{}903 (63)} & {\footnotesize{}900 (61)} & {\footnotesize{}900 (55)} & {\footnotesize{}900 (60)} & {\footnotesize{}899 (60)}\tabularnewline
\hline 
{\footnotesize{}9} & {\footnotesize{}879 (54)} & {\footnotesize{}879 (53)} & {\footnotesize{}882 (70)} & {\footnotesize{}876 (64)} & {\footnotesize{}879 (51)} & {\footnotesize{}882 (62)} & {\footnotesize{}880 (62)}\tabularnewline
\hline 
{\footnotesize{}8} & {\footnotesize{}798 (53)} & {\footnotesize{}798 (54)} & {\footnotesize{}801 (61)} & {\footnotesize{}798 (59)} & {\footnotesize{}798 (53)} & {\footnotesize{}798 (58)} & {\footnotesize{}796 (59)}\tabularnewline
\hline 
{\footnotesize{}7} & {\footnotesize{}654 (53)} & {\footnotesize{}654 (57)} & {\footnotesize{}663 (57)} & {\footnotesize{}657 (64)} & {\footnotesize{}654 (54)} & {\footnotesize{}657 (53)} & {\footnotesize{}652 (50)}\tabularnewline
\hline 
{\footnotesize{}6} & {\footnotesize{}621 (53)} & {\footnotesize{}621 (52)} & {\footnotesize{}621 (59)} & {\footnotesize{}621 (57)} & {\footnotesize{}621 (52)} & {\footnotesize{}621 (57)} & {\footnotesize{}619 (58)}\tabularnewline
\hline 
{\footnotesize{}5} & {\footnotesize{}489 (56)} & {\footnotesize{}492 (56)} & {\footnotesize{}492 (65)} & {\footnotesize{}498 (63)} & {\footnotesize{}489 (57)} & {\footnotesize{}489 (59)} & {\footnotesize{}491 (65)}\tabularnewline
\hline 
{\footnotesize{}4} & {\footnotesize{}456 (59)} & {\footnotesize{}456 (59)} & {\footnotesize{}456 (67)} & {\footnotesize{}456 (68)} & {\footnotesize{}456 (60)} & {\footnotesize{}453 (65)} & {\footnotesize{}454 (72)}\tabularnewline
\hline 
{\footnotesize{}3} & {\footnotesize{}252 (50)} & {\footnotesize{}252 (49)} & {\footnotesize{}252 (56)} & {\footnotesize{}252 (53)} & {\footnotesize{}252 (50)} & {\footnotesize{}249 (56)} & {\footnotesize{}254 (62)}\tabularnewline
\hline 
{\footnotesize{}2} & {\footnotesize{}201 (75)} & {\footnotesize{}201 (83) } & {\footnotesize{}204 (68)} & {\footnotesize{}204 (105)} & {\footnotesize{}201 (76)} & {\footnotesize{}198 (75)} & {\footnotesize{}193 (87)}\tabularnewline
\hline 
{\footnotesize{}1} & {\footnotesize{}78 (54)} & {\footnotesize{}78 (54)} & {\footnotesize{}81 (65)} & {\footnotesize{}78 (56)} & {\footnotesize{}78 (55)} & {\footnotesize{}81 (59)} & {\footnotesize{}79 (53)}\tabularnewline
\hline 
{\scriptsize{}ZPE} & {\footnotesize{}17164 (46)} & {\footnotesize{}17188 (55)} & {\footnotesize{}17185 (62)} & {\footnotesize{}17230 (57)} & {\footnotesize{}17161 (55)} & {\footnotesize{}17146 (58)} & {\footnotesize{}17046 (63)}\tabularnewline
\hline 
{\footnotesize{}MAE} & {\footnotesize{}-} & {\footnotesize{}2.1} & {\footnotesize{}3.8} & {\footnotesize{}4.7} & {\footnotesize{}0.4} & {\footnotesize{}3.8} & {\footnotesize{}8.8}\tabularnewline
\hline 
{\footnotesize{}MFWHM} & {\footnotesize{}57} & {\footnotesize{}59} & {\footnotesize{}62} & {\footnotesize{}63} & {\footnotesize{}58} & {\footnotesize{}61} & {\footnotesize{}66}\tabularnewline
\hline 
\end{tabular}
\end{table}

All the results presented in Table \ref{tab-DB-Gly} have excellent
accuracy with respect to the benchmark semiclassical calculation based
on 2500 \emph{ab initio} Hessians. FWHM data further demonstrate the
reliability of the various approximations. Due to the length of the
dynamics ($\approx$ 0.6 ps), the lower bound to the FWHM values of
the Fourier transform signal is about 30 cm\textsuperscript{-1}.
Results show that most of them are actually in the 50-60 cm\textsuperscript{-1}
range and occasionally larger. The main reasons for such an enlargement,
which apply also to the case of methane, lie in nearby states contributing
to the power spectrum and to spurious rotations due to the adoption
for SC calculations of a normal mode reference frame based on the
equilibrium geometry. Both the database approach and the HU Bofill
scheme are accurate, but DBH appears more stable and deteriorates
in accuracy more slowly as fewer Hessian calculations are performed
\emph{ab initio}. To better appreciate this point and the comparison
among the two approaches, the ratio (R) between the maximum of 2500
Hessians and the number of those actually calculated \emph{ab initio}
is indicated in the header of DBH columns (DBH=R). More specifically,
data in column 3 of Table \ref{tab-DB-Gly} are based on 643 \emph{ab
initio} Hessians, the ones in column 4 on 328 Hessians, and those
in column 5 on 197 Hessians. The corresponding saving with respect
to the total simulation time (including generation of the dynamics)
of the all-Hessian simulations is equal to about 70\%, 82\%, and 88\%
respectively.

\subsection{N-Acetyl-L-Phenylalaninyl-L-Methionine Amide}

\begin{figure}[H]
\begin{centering}
\includegraphics{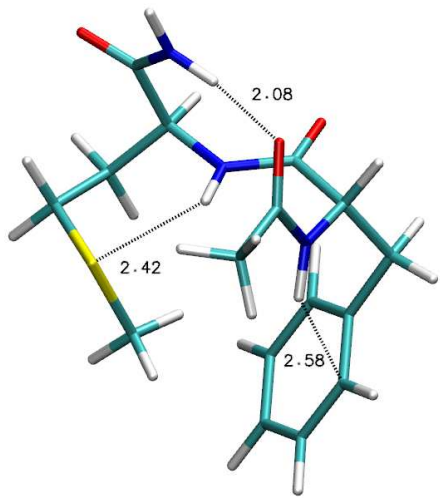}
\par\end{centering}
\caption{Global minimum of N-acetyl-L-phenylalaninyl-L-methionine amide at
DFT-B3LYP-D level of theory with 6-31G{*} basis set. Hydrogen bonds
are indicated by dashed lines.\label{46-atom-Global_Minimum}}
\end{figure}

Our final spectroscopic study was dedicated to the challenging 46-atom
N-acetyl-L-phenylalaninyl-L-methionine amide. The L-phenylalaninyl-L-methionine
dipeptide (Phe-Met) is a prototypical system of proteins in which
several hydrogen bonds govern the secondary and tertiary structures.
In particular, there are three different H-bonds which are crucial
in the stabilization of the folded structure: Two of them, an NH-{}-{}-$\pi$
in the phenylalaninyl and an NH-{}-{}-S in the methionine, are confined
within the lateral chain of the aminoacids, while the third one is
established between the NH\textsubscript{2} and C=O of the terminal
groups. Biswal et al. have studied recently the conformation and the
vibrational spectrum of the amide of this dipeptide capped with acetyl
(Ac-Phe-Met-NH$_{2}$). They presented quantum DFT-D calculations
together with gas-phase experiments, showing that the hydrogen bond
involving the sulfur atom has a strength similar to that of the intrabackbone
NH-{}-{}-O=C hydrogen bond.\citep{biswal2012strength}\\ Based on
the satisfactory results obtained for the previous systems, we decided
to employ the DBH approach. In fact, such a big and complex molecule
could not be investigated with reasonable computational effort by
means of the standard, all-Hessian SC procedure. DBH made the computational
overhead affordable, and we were able to undertake a study based on
\textit{ab initio} ``on-the-fly'' semiclassical dynamics at DFT-B3LYP-D
level of theory and 6-31G{*} basis set of the high energy quantum
fundamentals of vibration. These include the 2 stretches involving
the NH\textsubscript{2} group (sNH\textsubscript{2} and aNH\textsubscript{2})
plus the two NH stretches, labeled NH(I) and NH(II). We identified
the conformation illustrated in Fig. \ref{46-atom-Global_Minimum}
as the most stable one, and normal modes were determined for this
geometry.

As in the case of glycine, we performed \emph{ab initio} calculations
by means of the NWChem suite of codes,\citep{Valiev_DeJong_NWChem_2010}
but this time the MC-DC-SCIVR approach was employed. Following the
standard MC-DC-SCIVR recipe, we ran four simulations tailored on the
four normal modes of interest. The starting conditions were chosen
to be the equilibrium geometry for positions, while initial momenta
were assigned according to the harmonic ZPE with an additional quantum
of harmonic excitation for the mode under investigation. The reference
states were chosen as described in Eq. (\ref{eq:MCTASCIVR}) to enhance
the signal corresponding to the target mode. All the trajectories
were evolved for a total of about 0.6 ps. The subspace partition was
obtained in agreement with the Hessian criterion\citep{DiLiberto_Ceotto_Jacobiano_2018}
and monodimensional subspaces were determined for all four modes.
DBH was employed with $\rho=0.15$, which led to the construction
of databases of less than 300 Hessians.

\begin{figure}[H]
\begin{centering}
\includegraphics[scale=0.46]{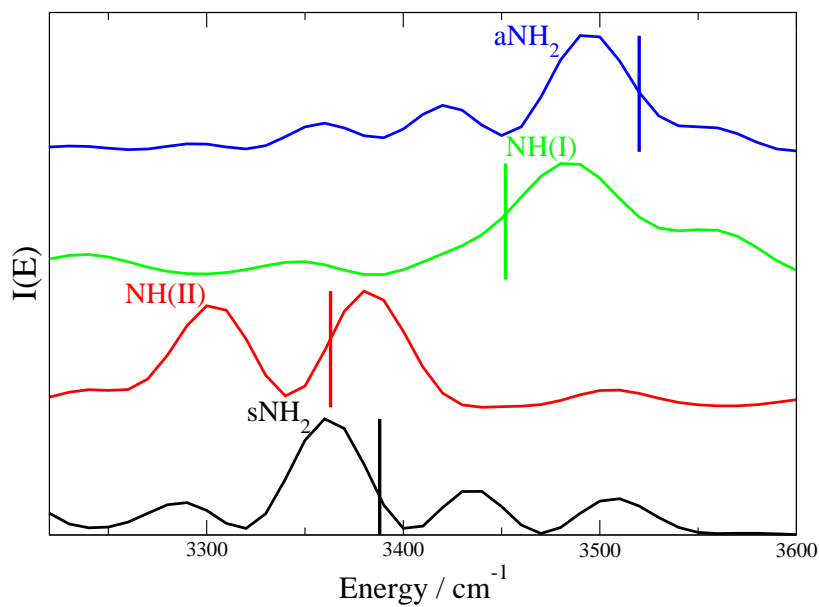}
\par\end{centering}
\caption{MC-DC-SCIVR power spectra for the 4 highest frequency fundamentals
of Ac-Phe-Met-NH$_{2}$. Vertical bars indicate experimental values.\label{46atom-refined} }
\end{figure}

The molecule is made of 132 vibrational degrees of freedom with nine
of them characterized by harmonic frequencies below 100 cm\textsuperscript{\textendash 1},
16 in the 1500-1800 cm\textsuperscript{-1} range, and 19 between
3000 and 3215 cm\textsuperscript{-1}. It is then clear that the density
of vibrational states is very high and diffuse in the 3300-3600 cm\textsuperscript{-1}
interval investigated in this work, which includes (in addition to
the four high frequency fundamentals) a large number of overtones
and combination bands. Consequently some bands, due to resonances
and dependent on the SC propagator (i.e. the trajectory) and the reference
state, rather than well isolated peaks are expected in the power spectra.
Furthermore, we found that for mode NH(I) the spectrum was very noisy
due to the many couplings to other modes, and a reliable frequency
estimate was not achievable. For this mode we ran a trajectory with
tailored reference state at the lower harmonic ZPE to help weaken
the couplings. It is known that such a simulation yields generally
less accurate (even if still acceptable) frequencies, but the outcome
was in this case proved to be satisfactory. Power spectra obtained
from the several simulations previously described are reported in
Fig. \ref{46atom-refined}, where labels indicating the vibrational
modes are written next to the relevant spectral signal. 

The corresponding semiclassical frequencies are reported in Tab. \ref{tab:FM_results}
together with experimental results and harmonic estimates. FWHM values
are given in parentheses. While harmonic values are substantially
off the mark, the discrepancy between semiclassical values and the
experiment is generally about 30 cm\textsuperscript{-1}. The distance
is bigger for the NH(II) mode, which features a double peak. The lower
frequency signal has been assigned to NH(II), while the other peak
has been assigned to the signal of the coupled sNH\textsubscript{2}
mode estimated in an approximate way using the SC propagator tailored
for the NH(II) mode. 

\begin{table}[H]
\caption{Experimental, MC-DC-SCIVR, and harmonic frequencies of Ace-Phe-Met-NH$_{2}$.
FWHM data are reported in parenthesis. All values are in cm\protect\textsuperscript{-1}.
\label{tab:FM_results}}

\renewcommand{\arraystretch}{1.4}
\centering{}%
\begin{tabular}{|c|c|c|c|}
\hline 
mode & Exp.\citep{biswal2012strength} & MC-DC SCIVR & Harm\tabularnewline
\hline 
\hline 
aNH\textsubscript{2} & 3520 & 3490 (47) & 3682\tabularnewline
\hline 
NH(I) & 3452 & 3480 (70) & 3607\tabularnewline
\hline 
NH(II) & 3363 & 3300 (45) & 3568\tabularnewline
\hline 
sNH\textsubscript{2} & 3388 & 3360 (45) & 3535\tabularnewline
\hline 
MAE & - & 37 & 167\tabularnewline
\hline 
\end{tabular}
\end{table}

\section{Summary and Conclusions\label{sec:Summary}}

We have introduced an innovative strategy to ease the calculation
of Hessian matrices required by semiclassical spectroscopy. It is
based on the idea that similar Hessians are likely to be derived from
similar geometries, and it consists in the dynamical construction
of a database of Hessian matrices and associated geometries. The other
Hessians needed are approximated by database records according to
a given threshold parameter accounting for the similarity of the instantaneous
geometry to those in the database. This novel approach is midway between
the single-Hessian approximation recently introduced in thawed Gaussian
semiclassical vibronic spectroscopy, which is generally not sufficiently
accurate for vibrational spectroscopy, and the basic full-Hessian
SC calculation, which is too computationally expensive for applications
to medium-large systems. 

For the same purpose some Hessian extrapolation schemes based on finite
differences have been developed. They have been shown to work efficiently
and satisfactorily for small molecules, but they depend on a linear
(monodimensional) approximation and their accuracy may deteriorate
in high dimensional applications. Furthermore, these Hessian update
schemes require new Hessian calculations rigorously at a constant
pace along the trajectory lacking the flexibility of the database
approach. 

We applied the DBH technique to methane, interfacing the method with
the Hessian Update Bofill scheme. This has permitted us to keep excellent
accuracy while calculating only about 6\% of the Hessians needed by
a regular simulation. However, the small dimensionality of the molecule
did not lead to a substantial speed up since the time needed by the
search algorithm offsets the time saved by the reduced number of Hessian
calculations. In this regard, the database approach has demonstrated
its power in applications to bigger molecules for which, in absence
of a precise analytical PES, \emph{ab initio} ``on-the-fly'' dynamics
is necessary. A first investigation of this kind involved glycine
and showed that DBH permits the preservation of very good accuracy,
reducing the number of Hessian matrices by a factor of about 13 and
total computational times by about 88\%. Hessian update schemes provided
good results too, but their accuracy deteriorated faster. The ultimate
demonstration of the upgrade provided by DBH is represented by the
final study of the high frequency power spectrum for Ac-Phe-Met-NH$_{2}$.
DBH has permitted the semiclassical analysis of this 46-atom system,
which otherwise would have not been attainable. 

Upon analysis of the results, we notice that the eigenvalues estimated
by DBH are losing accuracy much faster than frequencies which are
obtained as a difference between eigenvalues. This aspect does not
constitute a severe drawback when moving to high dimensional systems
for which the DC-SCIVR technique must be employed. In fact, ``divide
and conquer'' SCIVR approaches are themselves less accurate in estimating
ZPE energies compared to frequencies of vibration, which are the actual
target of investigation.

In perspective DBH will permit semiclassical studies of complex molecular
and supramolecular systems. In particular, the increased size of calculable
systems will enable the investigation of molecular solvation and molecular
adsorption on surfaces.
\begin{acknowledgments}
Authors acknowledge financial support from the European Research Council
(Grant Agreement No. (647107)\textemdash SEMICOMPLEX\textemdash ERC-
2014-CoG) under the European Union\textquoteright s Horizon 2020 research
and innovation programme, and from the Italian Ministery of Education,
University, and Research (MIUR) (FARE programme R16KN7XBRB- project
QURE). Part of the needed cpu time was provided by CINECA (Italian
Supercomputing Center) under ISCRAB project ``QUASP''. The research
was also supported in part by the US National Science Foundation under
Grant No. CNS-1526055.
\end{acknowledgments}

\bibliographystyle{aipnum4-1}
\bibliography{Biblio_RESUB}

\end{document}